# Quantum Multigrid Algorithm for Finite Element Problems

Osama Muhammad Raisuddin[1]*, Suvranu De[2]

*raisuo@rpi.edu

[1] *Mechanical, Aerospace, and Nuclear Engineering, Rensselaer Polytechnic Institute, 110 8th St. Troy, NY, 12180, USA*

[2] *Florida Agricultural and Mechanical University-Florida State University College of Engineering, 2525 Pottsdamer St., Tallahassee, FL 32310, USA*

**Keywords**: gate-based quantum computing; iterative methods; finite element; exponential advantage; multigrid method

## Abstract

Quantum linear system algorithms (QLSAs) can provide exponential speedups for the solution of linear systems, but the growth of the condition number for finite element problems can eliminate the exponential speedup. QLSAs are also incapable of using an initial guess of a solution to improve upon it. To circumvent these issues, we present a Quantum Multigrid Algorithm (qMG) for the iterative solution of linear systems by applying the sequence of multigrid operations on a quantum state. Given an initial guess with error $\epsilon_0$, qMG can produce a vector encoding the entire sequence of multigrid iterates with the final iterate having a relative error $\tilde{\epsilon} = \epsilon/\epsilon_0$, as a subspace of the final quantum state, with exponential advantage in $O\left(\text{poly}\log\frac{N}{\tilde{\epsilon}}\right)$ time using $O\left(\text{poly}\log\frac{N}{\tilde{\epsilon}}\right)$ qubits. Although extracting the final iterate from the sequence is efficient, extracting the sequence of iterates from the final quantum state can be inefficient. We provide an analysis of the complexity of the method along with numerical analysis.

## 1. Introduction

The solution of physical problems discretized using the finite element method requires solving a system of linear equations. Quantum linear system algorithms exhibit an exponentially improved scaling, $O(\log N)$ in the number of unknowns $N$ in comparison to classical algorithms which scale as $O(N)$ at best (Harrow et al., 2009). As an example, for a general indefinite system the conjugate gradient algorithm scales as $O(N\kappa\log(1/\epsilon))$, where $\kappa$ is the condition number of the system and $\epsilon$ is the desired precision. However, the quantum signal processing algorithm scales as $O(\kappa \text{ poly}\log(N\kappa/\epsilon))$ (Martyn et al., 2021) for the same linear system of equations. For quantum linear system algorithms to provide an overall exponential speedup, $\kappa$ must scale as $O(\text{poly}\log N)$ or better. This prevents a quantum speedup from direct application of QLSAs to finite element problems since $\kappa = O(N^2)$ in the worst case (Montanaro & Pallister, 2016). In this paper, we propose a multigrid algorithm to improve the solution complexity of finite element problems on quantum computers.

Quantum computing is an emerging computational paradigm with the potential to solve problems considered intractable using classical computing. The properties of superposition, entanglement and interference in exponentially large state spaces set quantum computers apart from classical computers. Unlike classical bits, quantum computers use quantum bits or 'qubits' to represent information. A quantum memory register is represented as an exponentially large vector in $\mathcal{H}^{2^n}$ as the state of $n$-qubits. The state of the qubits is represented as a continuum of superposition of basis vectors in $\mathcal{H}^{2^n}$. The state space of qubits includes 'entangled' states, which makes the state of qubits correlated to each other. Unwanted states in the superposition can be cancelled out using interference. Quantum computers were originally conceived to simulate quantum mechanical systems (Feynman, 1986). However, in recent decades other speedups using quantum computers have been discovered, including factoring large integers (Shor, 1994). Algorithms that exhibit great potential for scientific computing are quantum algorithms for linear systems of equations (Harrow et al., 2009), systems of ordinary differential equations (Berry, 2014a), and partial differential equations (Childs et al., 2020a), all of which have exponentially improved scaling.

Gate-based quantum computing (Nielsen & Chuang, 2011) and quantum annealing (Finnila et al., 1994) are the two prevalent quantum computing architectures. Quantum annealing is a continuous operation with the goal of preparing and measuring the ground state of a Hamiltonian. This is achieved by a continuous transition of qubits in a known ground state of an initial Hamiltonian to the unknown ground state of the final Hamiltonian. An iterative quantum annealing approach for finite element problems has been provided by (Raisuddin & De, 2022). Current-generation Noisy Intermediate-Scale Quantum (NISQ) quantum annealers show empirical quantum advantage over classical simulated annealing for some problems (King et al., 2015) but the only known provable speedup is an exponential speedup for the glued trees problem (Muthukrishnan et al., 2019).

This paper focuses on the gate-based quantum computing model, which represents a quantum 'circuit' consisting of quantum 'gates' applied to qubits in quantum registers. The output of a quantum algorithm can be in the form of classical measurements of the qubits or a quantum state. In gate-based quantum computers, data is typically represented using amplitude encoding (Weigold et al., 2021). The entries of a vector $\boldsymbol{u} \in \mathbb{C}^{2^n}$ are mapped to the probability amplitudes of a quantum state of $n$ qubits $|\boldsymbol{u}\rangle$. The quantum state is normalized using the inner product $|\langle\boldsymbol{u}|\boldsymbol{u}\rangle|_2^2 = 1$ due to the Born rule (Born, 1926; Landsman, 2009). An alternative data representation scheme is basis encoding (Weigold et al., 2021), which represents classical strings of $n$ bits as quantum states of $n$ qubits.

Access to problem parameters can be provided in the form of quantum oracles (Childs et al., 2017; Martyn et al., 2021). For a linear system of equations, an oracle may be used to access the entries of a matrix or prepare a quantum state $|\boldsymbol{b}\rangle$ corresponding to a vector $\boldsymbol{b}$. A powerful method of accessing matrix data in quantum computing is using block-encoded unitaries of matrices (Chakraborty et al., 2019). Block-encoded unitaries are used in qubitization methods, which are the basis for many optimal or near-optimal versions of quantum algorithms (Martyn et al., 2021).

A quantum linear system algorithm (QLSA) takes an amplitude-encoded quantum state $|\boldsymbol{b}\rangle$ as an input and outputs a state $|\boldsymbol{x}\rangle$ proportional to the solution of a $d$-sparse system of linear equations $\boldsymbol{Ax} = \boldsymbol{b}$, where $\boldsymbol{A} \in \mathbb{C}^{N\times N}$, $\boldsymbol{x}, \boldsymbol{b} \in \mathbb{C}^N$, and $d$ denotes the maximum number of non-zero entries in any row or column. The first QLSA was the HHL algorithm developed by (Harrow et al., 2009) with complexity $O(\log(N)\, d^2\kappa^2/\epsilon\,)$. The complexity of QLSAs has improved since then with a complexity of $O(d\kappa \operatorname{poly}\log(d\kappa N/\epsilon))$ using a sparse matrix query model and linear combination of unitary (LCU) techniques (Childs et al., 2017), which can be improved to $O(\kappa \operatorname{poly}\log(\kappa N/\epsilon))$ using a block-encoding query model combined with quantum signal processing (QSP) method. QLSAs can also deal with singular systems of equations if $\boldsymbol{b}$ lies in the null space of $\boldsymbol{A}$.

LCU and QSP QLSAs use a polynomial $P(\boldsymbol{A}) \approx \boldsymbol{A}^{-1}$ over the interval $[-1, -1/\kappa] \cup [1/\kappa, 1]$, where $\boldsymbol{A}$ is normalized s.t. $\|\boldsymbol{A}\| \leq 1$ s.t. $\max \|P(\boldsymbol{A}) - \boldsymbol{A}^{-1}\| \leq \epsilon$. The LCU method applies the individual components of the Chebyshev polynomial directly, which requires an overhead of several ancilla qubits due to 'select' and 'prepare' operations. The QSP method uses qubitization techniques (Martyn et al., 2021) to apply the polynomial using a QSP circuit. A QSP circuit for a particular problem requires a QSP angle sequence equivalent to the desired polynomial (Martyn et al., 2021). The QSPPACK (Dong et al., 2021) and PyQSP (Martyn et al., 2021) libraries provides angle sequences for desired polynomials. The degree of the Chebyshev polynomial is equal to the number of QSP rotation angles and scales $\propto \kappa \log 1/\epsilon$. As a consequence, the depth of the quantum circuit is linear in $\kappa$. The QSP angles can be increasingly difficult to calculate for higher order polynomials due to loss of numerical precision. The QSP method has been generalized to accommodate all possible rotations (Motlagh & Wiebe, 2023) but an implementation of the method is not available. Recent work has also shown that approximating only over the interval $[1/\kappa, 1]$ allows solution of certain classes of symmetric positive-definite systems with $O\left(\sqrt{\kappa} \operatorname{poly}\log(\kappa N/\epsilon)\right)$ scaling instead of the linear scaling in $\kappa$ (Orsucci & Dunjko, 2021), which matches classical optimal scaling in $\kappa$.

Quantum algorithms for partial differential equations have been studied by (Childs et al., 2020b) and (Arrazola et al., 2019). (Childs et al., 2020b) solves the Laplace and second-order elliptic problems on square or rectangular domains with regularly spaced grid points using finite difference or spectral methods. (Arrazola et al., 2019) inverts polynomial differential operators instead. The first preconditioned quantum algorithm for the finite element problems was proposed by (Clader et al., 2013), using an SPAI (Sparse Approximate Inverse) left-preconditioner. However, explicit implementation details are not provided. (Montanaro & Pallister, 2016) point out in their analysis that while producing a quantum state proportional to the solution of partial differential equations using the finite element method can be exponentially efficient, extracting information about the output can eliminate the exponential speedups, but polynomial speedups are possible for problems with large higher-order derivatives and spatial dimensions. A comprehensive review and theory of quantum algorithms for differential equations can be found in (An et al., 2022).

Iterative solution methods are foundational for efficient solutions of problems in mechanics. Optimal iterative methods like the multigrid method can scale as $O(N)$ (Briggs et al., 2000) compared to the $O(N^3)$ scaling of direct methods (Trefethen & Bau, 1997). However, iterative methods have not been explored thoroughly in the context of quantum computing. A limiting factor is calculating inner products since it entails measuring a quantum state. However, relaxation or smoothing methods for positive-definite systems do not require calculating inner products (Axelsson, 1994), making them a viable option. (Raisuddin & De, 2023) provides an efficient relaxation method for positive-definite linear systems which scales exponentially better.

In this paper we present qMG, the first multigrid approach on quantum computers by encoding all the linear operations of the multigrid method into a larger matrix and encoding all multigrid iterates in a larger vector. The exponentially large state space of qubits allows efficient storage and manipulation of the larger vector encoding the multigrid iterates. Using block-encoding of the larger matrices, we show that the sequence of multigrid iterates can be produced in a subspace of the final vector with exponential advantage. The quantum state corresponding to the desired final iterate is prepared by performing measurements of ancillary qubits. This approach differs from (Raisuddin & De, 2023) since it uses a time-marching strategy used by (An et al., 2022; Fang et al., 2023) to apply linear operations instead of the technique based on solving a block-lower-triangular linear system developed by (Berry, 2014b). The aim of the qMG algorithm is to prepare a quantum state encoding the output of the multigrid method. The quantum state can then be used to obtain scalar properties of the solution using quantum measurements.

The paper is organized as follows: Section 2 introduces the solution of finite element problems using the multigrid method. Section 3 introduces our notation and method for encoding multigrid operations on quantum computers, along with the amplitude encoding of multigrid iterates as a quantum state. In Section 4 we introduce block-encodings of matrices into unitaries. In Section 5 we provide the quantum multigrid algorithm and analyze its overall complexity along with numerical results. Finally, we conclude in Section 6 with a discussion.

# 2. Finite element problem

We consider the following linear finite element problem defined on the discretized domain $\Omega_{\mathrm{h}} \in \mathbb{R}^{\mathrm{d}}, \mathrm{d} \in \{1,2,3\}$ with boundary $\Gamma_{\mathrm{h}}$

*Find* $\boldsymbol{u} \in \mathbb{R}^{\mathrm{N}}$ *such that*

$$\boldsymbol{\Psi}(\boldsymbol{u}) = \mathbf{A}\boldsymbol{u} - \mathbf{b} = \mathbf{0} \tag{1}$$

subject to $\boldsymbol{u} = \boldsymbol{u}_g$ on $\Gamma_{\mathrm{g}}$

where $\boldsymbol{u}$ is a vector of nodal unknowns, $\boldsymbol{A} \in \mathbb{R}^{\mathrm{N}\times\mathrm{N}}$ is the system matrix, $\boldsymbol{b} \in \mathbb{R}^{\mathrm{N}}$ is the forcing function, $N$ is the number of degrees of freedom of the discretized problem and $\Gamma_{\mathrm{g}}$ is the Dirichlet boundary. The system in equation (1) arises from the discretization of the weak form and incorporates the Neumann boundary conditions.

## 2.1. Multigrid method

We denote an approximation to the exact solution $\boldsymbol{u}$ as $\boldsymbol{v}$.

The multigrid method uses a series of $\mathcal{L} + 1$ discretized approximations $\{\boldsymbol{v}^L \mid L \in \mathbb{W} \leq \mathcal{L}\}$ on increasingly coarse grids, along with inter-grid transfer operators to transfer the discretized solution from the finer grids to coarser grids, and vice versa, known as prolongation and restriction operators $\boldsymbol{I}_L^{L+1}$ and $\boldsymbol{I}_{L+1}^{L}$ , respectively.

We denote the finest and coarsest grids by $L = 0$ and $L = \mathcal{L}$, respectively. One choice of transfer operators satisfies the Galerkin property:

$$\overset{L+1}{\boldsymbol{A}} = \boldsymbol{I}_L^{L+1}\overset{L}{\boldsymbol{A}}\boldsymbol{I}_{L+1}^{L} \tag{2}$$

The multigrid method uses iterative techniques to obtain corrections to fine-grid solutions at coarse grid levels, and the corrections are then transferred back to the fine grid level.

To achieve this, first $\nu_1$ relaxation or pre-smoothing steps are performed, typically using Gauss-Seidel or Richardson iterations of the general form:

$$\boldsymbol{v}^{L,i+1} = \boldsymbol{R}_i \boldsymbol{v}^{L,i} + \omega_i \boldsymbol{f} \tag{3}$$

where $\boldsymbol{R}_i = \boldsymbol{I} - \omega_i \boldsymbol{A}$

Here, we consider stationary linear iterations, i.e., $\omega = \omega_i$, and absorb the coefficient $\omega$ into $\boldsymbol{A}$ and $\boldsymbol{f}$. The problem of finding a correction to the pre-smoothed solution at grid level $L$ is posed using the residual equation

$$\boldsymbol{r}^L = \boldsymbol{f}^L - \overset{L}{\boldsymbol{A}} \boldsymbol{v}^L \tag{4}$$

The residual at the fine grid level $L$ is then prolonged to the coarser grid $L + 1$ using the inter-grid transfer operator

$$\boldsymbol{f}^{L+1} = \boldsymbol{I}_L^{L+1} \boldsymbol{r}^L \tag{5}$$

The pre-smoothing steps, residual equation, and transfer operations are repeated till the coarsest grid level is reached. At the coarsest grid level, either a direct solver or additional smoothing steps can be used to obtain the correction. The finer grid solutions are then corrected using the restriction operation

$$\boldsymbol{v}^{L,i+2} = \boldsymbol{I}_{L+1}^{L} \boldsymbol{v}^{L+1} + \boldsymbol{v}^{L,i+1}$$

After the restriction operation, $\nu_2$ post-smoothing steps are performed using additional iterations before restricting the solution to the next finer grid till the finest grid level is reached.

This sequence of operations from the finest grid level to the coarsest grid level and back to the finest grid level is denoted as a V-cycle. One may perform several V-cycles till the error is reduced to the acceptable criterion.

We summarize a V-cycle as a recursive algorithm in the following table:

Algorithm 1. Multigrid V-cycle: $\boldsymbol{v}^L \leftarrow V^L(\boldsymbol{v}^L, \boldsymbol{f}^L)$

**Input**: Initial guess $\boldsymbol{v}^{\mathcal{L}}$, $\{\boldsymbol{I}_L^{L+1}, \boldsymbol{I}_{L+1}^{L} | L \in \mathbb{W} < \mathcal{L}\}$, $\left\{\overset{L}{\boldsymbol{A}} | L \in \mathbb{W} < \mathcal{L}\right\}$, $\boldsymbol{f}^{\mathcal{L}}, \nu_1, \nu_2$

**Output**: $\boldsymbol{v}^{\mathcal{L}}$

1. $\nu_1$ pre-smoothing steps on $\overset{L}{\boldsymbol{A}} \boldsymbol{u}^L = \boldsymbol{f}^L$ with initial guess $\boldsymbol{v}^L$
2. **If** $L = \mathcal{L}$, go to 4.
   **Else**
   - $\boldsymbol{f}^{L+1} \leftarrow \boldsymbol{I}_L^{L+1} \left(\boldsymbol{f}^L - \overset{L}{\boldsymbol{A}} \boldsymbol{v}^L\right)$
   - $\boldsymbol{v}^{L+1} \leftarrow \boldsymbol{0}$
   - $\boldsymbol{v}^{L+1} \leftarrow V^{L+1}(\boldsymbol{v}^{L+1}, \boldsymbol{f}^{L+1})$
3. Correct $\boldsymbol{v}^L \leftarrow \boldsymbol{v}^L + \boldsymbol{I}_{L+1}^{L} \boldsymbol{v}^{L+1}$
4. Relax $\nu_2$ times on $\overset{L}{\boldsymbol{A}} \boldsymbol{u}^L = \boldsymbol{f}^L$

## 3. Multigrid operations on quantum computers

In this section, we provide a method to encode the operations of the multigrid algorithm into a quantum algorithm. We use block-encoded matrices to encode the sequence of multigrid operations.

## 3.1. Notation and indexing

We first define our notation for the multigrid V-cycles:

$\mathcal{V}+1$: Total number of V-cycles

$V$: V-cycle number $\qquad 0 \le V \le \mathcal{V} \in \mathbb{W}$

$\mathcal{L}+1$: Total number of grid levels

$L$: Grid level $\qquad 0 \le L \le \mathcal{L} \in \mathbb{W}$

$\nu - 1$: Number of pre- and post-smoothing steps

$v$: iterate number $\qquad 0 \le v \le 2\nu - 1 \in \mathbb{W}$

$v = 0$ denotes the initial guess or approximation to $\boldsymbol{u}$.

$1 \le v \le \nu - 1$ denotes the pre-smoothing iterates.

$v = \nu$ denotes the restricted solution.

$\nu + 1 \le v \le 2\nu - 1$ denotes the post-smoothing iterates.

We now define our notation to address the multigrid iterates and residuals across V-cycles. We use $\boldsymbol{v}_{V,L,v}$ to denote the $v^{th}$ multigrid iterate for V-cycle $V$ and grid level $L$. The residual for V-cycle $V$ at grid level $L$ before prolongation to grid level $L+1$ is denoted as $\boldsymbol{r}_V^{\substack{L \\ L+1}}$, and after prolongation to grid level $L+1$ is denoted as $\boldsymbol{r}_V^{L+1}$.

We use the notation of the multigrid iterates and residuals to combine all iterates into the vector $\boldsymbol{x}$, defined in Section 3.2.

For convenient block-indexing of the iterates in the vector $\boldsymbol{x}$, we define the following:

$T_V = 2(\mathcal{L}+1)\nu + 2\mathcal{L} - 1$: Number of blocks in a V-cycle (6)

$T = \mathcal{V}T_V + 2\mathcal{L}\nu + 2\mathcal{L} + 2\nu - 1$: Total number of blocks for multigrid operations (7)

$iPre(V, L, v) = V \cdot T_V + L(\nu + 2) + v$: Indexing for pre-smoothing steps (8)

$iPost(V, L, v) = VT_V + 2\mathcal{L}(\nu + 1) - L\nu + v$: Indexing for post-smoothing steps (9)

We also use the bra-ket or Dirac notation to denote quantum registers and operations on quantum computers. A quantum register of qubits is described as a 'ket' vector $|\psi\rangle$ where $\psi$ is a descriptor of the state of the register. In this paper, we use scalars $|i\rangle$ where $i \in \mathbb{W}$ to denote a quantum register in the $i^{th}$ basis state, equivalent to the $i^{th}$ standard basis vector $\boldsymbol{e}_i$. A vector $|\boldsymbol{x}\rangle$ where $\boldsymbol{x} \in \mathbb{R}^N$ denotes a quantum state that encodes $\boldsymbol{x}$ in the probability amplitudes of its basis states. A 'bra' $\langle\psi|$ is the conjugate transpose of a ket $|\psi\rangle$. $\langle\psi|\phi\rangle$ denotes an inner product of a bra $\langle\psi|$ and ket $|\phi\rangle$. A quantum state is always normalized to satisfy $|\langle\psi|\psi\rangle|^2 = 1$, however we occasionally omit the overall normalization constants of quantum states for brevity and clarity without any loss of generalization.

## 3.2. Block-encoding of multigrid iterates and residuals

We can define the multigrid iterates of the pre-smoothing steps as

$$\boldsymbol{pre}_{V,L} = \left(\boldsymbol{v}_{V,L,0}^T, \boldsymbol{v}_{V,L,1}^T, \dots, \boldsymbol{v}_{V,L,\nu-1}^T\right) \tag{10}$$

Similarly, the residuals before and after prolongation are denoted as

$$\boldsymbol{res}_{V,L} = \left( \boldsymbol{r}_V^{\,L\ T}_{\,L+1}, \boldsymbol{r}_V^{\,L+1^T} \right) \tag{11}$$

Finally, we define the corrected iterates and the subsequent post-smoothing steps as

$$\boldsymbol{post}_{V,L} = \left(\boldsymbol{v}_{V,L,\nu}^T, \boldsymbol{v}_{V,L,\nu+1}^T, \dots, \boldsymbol{v}_{V,L,2\nu-1}^T\right) \tag{12}$$

Note that for subsequent V-cycles the initial guess is the output/final iterate of the previous V-cycle as

$$\boldsymbol{v}_{V,\mathcal{L},0} = \boldsymbol{v}_{V-1,\mathcal{L},2\nu-1} \tag{13}$$

Combining all the iterates for a single V-cycle, we define

$$\boldsymbol{cycle}_V = \left(\boldsymbol{pre}_{V,0}, \boldsymbol{res}_{V,0}, \boldsymbol{pre}_{V,1}, \boldsymbol{res}_{V,1}, \dots, \boldsymbol{pre}_{V,\mathcal{L}-1}, \boldsymbol{res}_{V,\mathcal{L}-1}, \boldsymbol{pre}_{V,\mathcal{L}}, \boldsymbol{post}_{V,\mathcal{L}}, \boldsymbol{post}_{V,\mathcal{L}-1}, \dots \boldsymbol{post}_{V,0}\right) \tag{14}$$

We combine all the iterates for $\mathcal{V}$ V-cycles into the vector $\boldsymbol{m}$:

$$\boldsymbol{m} = \left((\boldsymbol{cycle}_0)_{:-1}, (\boldsymbol{cycle}_1)_{:-1}, \dots, (\boldsymbol{cycle}_{\mathcal{V}-1})_{:-1}, \boldsymbol{cycle}_{\mathcal{V}}\right) \tag{15}$$

Where $(\boldsymbol{cycle}_V)_{:-1}$ denotes truncation to exclude the last vector block $\boldsymbol{v}_{V,L,2\nu-1}^T$ since $\boldsymbol{v}_{V,\mathcal{L},0} = \boldsymbol{v}_{V-1,\mathcal{L},2\nu-1}$. We also define the vector $\boldsymbol{c}$, which contains $c$ copies of the final iterate $\boldsymbol{v}_{\mathcal{V},0,2\nu-1}$

$$\boldsymbol{c} = \left(\boldsymbol{v}_{\mathcal{V},0,2\nu-1}^T, \dots, \boldsymbol{v}_{\mathcal{V},0,2\nu-1}^T\right) \tag{16}$$

Finally, we define the vector $\boldsymbol{x}$, which the qMG algorithm produces as a quantum state:

$$\boldsymbol{x} = (\boldsymbol{m}, \boldsymbol{c})^T \tag{17}$$

The vector $\boldsymbol{x}$ contains the multigrid iterates ($\boldsymbol{m}$) along with $c$ copies of the final iterate ($\boldsymbol{c}$), which are used to boost the success probability of obtaining the final iterate, which is discussed in Section 5. The iterates and copies in $\boldsymbol{x}$ are padded with zeros to maintain a consistent block size with the finest grid vector for convenient indexing.

To track the progress of the algorithm across various operations, we additionally define the following notation for a block-slice of the vector by defining an indexing register $|i\rangle$ and a work register $|\boldsymbol{x}_i\rangle$:

$$|\boldsymbol{x}_{0:j}\rangle = \sum_{0 \le i \le j} |i\rangle\langle i| \otimes |\boldsymbol{x}_i\rangle \tag{18}$$

where $|i\rangle$ is the $i^{th}$ basis vector of the indexing register. We occasionally omit normalization of quantum states for clarity and brevity, with no loss of generalization.

The goal of the quantum multigrid algorithm is to start with an initial state $|\boldsymbol{x}_{in}\rangle$ and prepare the quantum state

$$|\boldsymbol{x}_{out}\rangle = |\boldsymbol{x}_j\rangle = \frac{\boldsymbol{v}_{\mathcal{V},0,2\nu-1}}{\|\boldsymbol{v}_{\mathcal{V},0,2\nu-1}\|}. \tag{19}$$

This is achieved when the indexing register is measured in the state $|j\rangle$.

## 3.3. Block-encoding of multigrid operations

In this section, we define the block-matrices used to prepare the quantum state $|\boldsymbol{x}\rangle$. Since all operations in the multigrid method are linear, they can be defined as block-linear operations as either matrix multiplications or through back-substitution. In this paper, implement the block-linear operations as matrix multiplications.

The pre-smoothing or post-smoothing steps may be performed by $\nu$ applications of a block-encoded iteration matrix

$$\begin{pmatrix} \boldsymbol{I} & & & & & & \\ & \ddots & & & & & \\ & & \boldsymbol{I} & & & & \\ & & & \boldsymbol{I} & & & \\ & & & \boldsymbol{R} & \boldsymbol{I} & & \\ & & & & & \boldsymbol{I} & \\ & & & & & & \ddots \\ & & & & & & & \boldsymbol{I} \end{pmatrix} \begin{pmatrix} \boldsymbol{v}_{V,L,0} \\ \vdots \\ \boldsymbol{v}_{V,L,\nu-2} \\ \boldsymbol{v}_{V,L,\nu-1} \\ \boldsymbol{f} \\ \boldsymbol{f} \\ \vdots \\ \boldsymbol{f} \end{pmatrix} = \begin{pmatrix} \boldsymbol{v}_{V,L,0} \\ \vdots \\ \boldsymbol{v}_{V,L,\nu-2} \\ \boldsymbol{v}_{V,L,\nu-1} \\ \boldsymbol{v}_{V,L,\nu} \\ \boldsymbol{f} \\ \vdots \\ \boldsymbol{f} \end{pmatrix} \tag{20}$$

Similarly, the residual equation, prolongation and restriction operations can be applied as block matrix multiplication:

$$\begin{pmatrix} \boldsymbol{v}_{V,L,\nu-1} \\ {}^{L}_{L+1}\boldsymbol{r} \end{pmatrix} = \begin{pmatrix} \boldsymbol{I} & \\ -\boldsymbol{A} & \boldsymbol{I} \end{pmatrix} \begin{pmatrix} \boldsymbol{v}_{V,L,\nu-1} \\ \boldsymbol{f} \end{pmatrix} \tag{21}$$

$$\begin{pmatrix} {}^{L}_{L+1}\boldsymbol{r} \\ {}^{L+1}\boldsymbol{r} \end{pmatrix} = \begin{pmatrix} \boldsymbol{I} & \\ \boldsymbol{I}_{L}^{L+1} & \boldsymbol{I} \end{pmatrix} \begin{pmatrix} {}^{L}_{L+1}\boldsymbol{r} \\ \boldsymbol{0} \end{pmatrix}, \tag{22}$$

$$\begin{pmatrix} \boldsymbol{v}_{V,L,\nu-1} \\ \boldsymbol{v}_{V,L+1,\nu-1} \\ \boldsymbol{v}_{v,L,\nu} \end{pmatrix} = \begin{pmatrix} \boldsymbol{I} & & \\ & \boldsymbol{I} & \\ \boldsymbol{I} & \boldsymbol{I}_{L+1}^{L} & \boldsymbol{I} \end{pmatrix} \begin{pmatrix} \boldsymbol{v}_{V,L,\nu-1} \\ \boldsymbol{v}_{V,L+1,\nu-1} \\ \boldsymbol{0} \end{pmatrix} \tag{23}$$

Finally, we define 'copy' operations to copy residuals to the appropriate blocks and also to boost the success probability of the algorithm by copying the final iterate to subsequent blocks (discussed in detail in Section 3.10) using matrix multiplications of the form:

$$\begin{pmatrix} \boldsymbol{I} & & & & & & \\ & \ddots & & & & & \\ & & \boldsymbol{I} & & & & \\ & & & \boldsymbol{I} & & & \\ & & & \boldsymbol{I} & \boldsymbol{I} & & \\ & & & & & \boldsymbol{I} & \\ & & & & & & \ddots \\ & & & & & & & \boldsymbol{I} \end{pmatrix} \begin{pmatrix} \boldsymbol{v}_{\mathcal{V},0,2\nu-1} \\ \vdots \\ \boldsymbol{v}_{\mathcal{V},0,2\nu-1} \\ \boldsymbol{v}_{\mathcal{V},0,2\nu-1} \\ \boldsymbol{0} \\ \boldsymbol{0} \\ \vdots \\ \boldsymbol{0} \end{pmatrix} = \begin{pmatrix} \boldsymbol{v}_{\mathcal{V},0,2\nu-1} \\ \vdots \\ \boldsymbol{v}_{\mathcal{V},0,2\nu-1} \\ \boldsymbol{v}_{\mathcal{V},0,2\nu-1} \\ \boldsymbol{v}_{\mathcal{V},0,2\nu-1} \\ \boldsymbol{0} \\ \vdots \\ \boldsymbol{0} \end{pmatrix} \tag{24}$$

### 3.4. Initial state

We begin our algorithm with a given initial guess $|\boldsymbol{v}_{0,0,0}\rangle$ and the forcing vector $\left|\boldsymbol{f}^{\mathcal{L}}\right\rangle$ as the input via. oracles $O_{\boldsymbol{v}}$ and $O_{\boldsymbol{f}}$. We then prepare the initial state

$$|\boldsymbol{x}_{in}\rangle = |0\rangle|\boldsymbol{x}_{0,0,0}\rangle + \sum_{\substack{0\le V\le \mathcal{V}\\ 1\le v\le \nu}}|iPre(V,0,v)\rangle|\boldsymbol{f}\rangle + \sum_{\substack{0\le V\le \mathcal{V}\\ \nu+1\le v\le 2\nu+1}}|iPost(V,0,v)\rangle|\boldsymbol{f}\rangle \quad (25)$$

Similar to the block-slicing for the final state, we define a block-slicing for the initial vector as:

$$\left|\boldsymbol{x}_{in_{i:T}}\right\rangle = \sum_{i<j\le T}|j\rangle\langle j| \otimes \left|\boldsymbol{x}_{in_j}\right\rangle \quad (26)$$

### 3.5. Pre-smoothing and residual equation

For the V-cycle $V$ and grid level $L$ we apply pre-smoothing steps and calculate the residual by applying the following matrix multiplications

$$\left|\boldsymbol{x}_{0:iPre(V,L,\nu)}\right\rangle + \left|\boldsymbol{x}_{in_{iPre(V,L,\nu):T}}\right\rangle + \sum_{1<l\le L}\left|\boldsymbol{r}_{V,l_{Post}}\right\rangle = \left[\prod_{1\le v\le \nu}\boldsymbol{Pre}_{V,L,v}\right]\left[\left|\boldsymbol{x}_{0:iPre(V,L,0)}\right\rangle + \left|\boldsymbol{x}_{in_{iPre(V,L,0):T}}\right\rangle + \left|\boldsymbol{r}_{V,L_{Pre}}\right\rangle + \sum_{1<l\le L}\left|\boldsymbol{r}_{V,l_{Post}}\right\rangle\right] \quad (27)$$

where

$$\boldsymbol{Pre}_{V,L,v} = |iPre(V,L,v)\rangle\langle iPre(V,L,v-1)| \otimes \left(\left(1-\delta_{\mathcal{L},L}\delta_{v,\nu}\right)\boldsymbol{R} + \delta_{\mathcal{L},L}\delta_{v,\nu}\boldsymbol{A}\right) + \sum_{0\le i\le T}|i\rangle\langle i| \otimes \boldsymbol{I} \quad (28)$$

and $\delta_{i,j}$ is the Kronecker delta.

The quantum state $\left|\boldsymbol{r}_{V,L_{Post}}\right\rangle$ is defined in Section 3.7.

Note that for the coarsest grid level, instead of computing the residual, an additional relaxation step is performed to initialize the post-smoothing steps.

### 3.6. Restriction

After the residual is calculated at a particular grid level, we restrict the residual to the coarse grid level by applying the following matrix multiplication:

$$\left|\boldsymbol{x}_{0:iPre(V,L,\nu)+1}\right\rangle + \left|\boldsymbol{x}_{in_{iPre(V,L,\nu)+1:T}}\right\rangle + \sum_{1<l\le L}\left|\boldsymbol{r}_{V,l_{Post}}\right\rangle = \boldsymbol{Res}_{V,L}\left[\left|\boldsymbol{x}_{0:iPre(V,L,\nu)}\right\rangle + \left|\boldsymbol{x}_{in_{iPre(V,L,\nu):T}}\right\rangle + \sum_{1<l\le L}\left|\boldsymbol{r}_{V,l_{Post}}\right\rangle\right] \quad (29)$$

where

$$\boldsymbol{Res}_{V,L} = |iPre(V,L,\nu)+1\rangle\langle iPre(V,L,\nu)| \otimes \boldsymbol{I}_L^{L+1} + \sum_{0\le i\le T}|i\rangle\langle i| \otimes \boldsymbol{I} \quad (30)$$

### 3.7. Residual copies

After the residual is restricted to the coarse grid, we 'copy' it to the requisite blocks, where it will be needed for smoothing operations, by applying the following sequence of matrix multiplications

$$\left|\boldsymbol{x}_{0:iPre(V,L,v)+1}\right\rangle + \left|\boldsymbol{x}_{in_{iPre(V,L,v)+1:T}}\right\rangle + \left|\boldsymbol{r}_{V,L+1_{Pre}}\right\rangle + \sum_{1<l\leq L+1}\left|\boldsymbol{r}_{V,l_{Post}}\right\rangle = \left[\prod_{v<v<2v}\boldsymbol{PostCopy}_{V,L,v}\right]\left[\prod_{1\leq v\leq v-1}\boldsymbol{PreCopy}_{V,L,v}\right]\left[\left|\boldsymbol{x}_{0:iPre(V,L,v)+1}\right\rangle + \left|\boldsymbol{x}_{in_{iPre(V,L,v)+1:T}}\right\rangle + \sum_{1<l\leq L+1}\left|\boldsymbol{r}_{V,l_{Post}}\right\rangle\right] \quad (31)$$

where

$$\boldsymbol{PreCopy}_{VL,v} = \left(1-\delta_{v,1}\right)|iPre(V,L+1,v)\rangle\langle iPre(V,L+1,v)-1| \otimes \boldsymbol{I} + \delta_{v,1}|iPre(V,L+1,v)\rangle\langle iPre(V,L,v)+1| \otimes \boldsymbol{I} \quad (32)$$

and

$$\boldsymbol{PostCopy}_{V,L,v} = \left(1-\delta_{v,v+1}\right)|iPost(V,L+1,v)\rangle\langle iPost(V,L+1,v)-1| \otimes \boldsymbol{I} + \delta_{v,v+1}|iPre(V,L+1,v)\rangle\langle iPost(V,L+1,v+1)| \otimes \boldsymbol{I} \quad (33)$$

and the residual copies are defined as

$$\left|\boldsymbol{r}_{V,L_{Pre}}\right\rangle = \sum_{1\leq v\leq v-1}|iPre(V,L,v)\rangle\left|\boldsymbol{r}_V^{L+1}\right\rangle \quad (34)$$

and

$$\left|\boldsymbol{r}_{V,L_{Post}}\right\rangle = \sum_{v+1\leq v\leq 2v-1}|iPost(V,L,v)\rangle\left|\boldsymbol{r}_V^{L+1}\right\rangle \quad (35)$$

The residuals for the post-smoothing steps come from the calculation and copying of residuals from the finer grid levels and will be used for the post-smoothing operations.

### 3.8. Post-smoothing

After all the pre-smoothing steps are completed at all grid levels, we start applying the post-smoothing operations, using a similar process as the pre-smoothing process, consuming the copies of the residual

$$\left|\boldsymbol{x}_{0:iPost(V,L,2v-1)}\right\rangle + \left|\boldsymbol{x}_{in_{iPost(V,L,2v-1):T}}\right\rangle + \sum_{1<l\leq L-1}\left|\boldsymbol{r}_{V,l_{Post}}\right\rangle = \left[\prod_{v<v<2v}\boldsymbol{Post}_{V,L,v}\right]\left[\left|\boldsymbol{x}_{0:iPost(V,L,v)}\right\rangle + \left|\boldsymbol{x}_{in_{iPost(V,L,v):T}}\right\rangle + \sum_{1<l\leq L}\left|\boldsymbol{r}_{V,l_{Post}}\right\rangle\right] \quad (36)$$

where

$$\boldsymbol{Post}_{V,L,v} = |iPost(V,L,v)\rangle\langle iPost(V,L,v-1)| \otimes \boldsymbol{R} + \sum_{0\leq i\leq T}|i\rangle\langle i| \otimes \boldsymbol{I} \quad (37)$$

### 3.9. Prolongation

Between the post-smoothing steps at various grid levels, we apply the prolongation operation by applying the following matrix multiplication:

$$\left|\boldsymbol{x}_{0:iPost(V,L-1,v)}\right\rangle + \left|\boldsymbol{x}_{in_{iPost(V,L-1,v):T}}\right\rangle + \sum_{1<l\leq L-1}\left|\boldsymbol{r}_{V,l_{Post}}\right\rangle = \boldsymbol{Pro}_{V,L}\left[\left|\boldsymbol{x}_{0:iPost(V,L,2v-1)}\right\rangle + \left|\boldsymbol{x}_{in_{iPost(V,L,2v-1):T}}\right\rangle + \sum_{1<l\leq L-1}\left|\boldsymbol{r}_{V,l_{Post}}\right\rangle\right] \quad (38)$$

where

$$\boldsymbol{Pro}_{V,L} = |iPost(V,L,v)+1\rangle\langle iPost(V,L,v-1)| \otimes \boldsymbol{I}_{L+1}^{L} + |iPost(V,L,v)+1\rangle\langle iPre(V,L,v-1)| \otimes \boldsymbol{I} + \sum_{0\leq i\leq T} |i\rangle\langle i| \otimes \boldsymbol{I}$$

### 3.10. Copies

After the final iterate of the final V-cycle has been calculated, it is copied into the subsequent empty vector blocks to boost the probability of isolating the final iterate when the indexing qubits are measured. This is performed by applying the following operations:

$$|\boldsymbol{x}_{0:T+c}\rangle = \left[\prod_{1\leq j\leq c} \boldsymbol{Copy}_j\right] |\boldsymbol{x}_{0:T}\rangle \qquad (39)$$

where

$$\boldsymbol{Copy}_j = |T+j\rangle\langle T+j-1| \otimes \boldsymbol{I} + \sum_{0\leq i\leq T} |i\rangle\langle i| \otimes \boldsymbol{I} \qquad (40)$$

# 4. Block encoded unitary matrices

Since all operators on quantum computers are unitary, except for measurement, non-unitary matrices such as the multigrid operations described in Section 3 cannot directly be multiplied with a quantum state. A powerful method for multiplying matrices, unitary or non-unitary, with a quantum state is to use block-encoded unitaries. An $(\alpha, a)$-block-encoding of a matrix $\boldsymbol{A}$ where $\|\boldsymbol{A}\| \leq \alpha$ can be defined as

$$\boldsymbol{U_A} = \begin{pmatrix} \boldsymbol{A}/\alpha & * \\ * & * \end{pmatrix} \qquad (41)$$

where $\boldsymbol{A} = \alpha(\langle 0^a| \otimes I)\boldsymbol{U_A}(I \otimes |0^a\rangle)$ and $\alpha$ is the subnormalization factor. A block encoding of a matrix may be created using the Linear Combination of Unitaries method and may use several ancilla qubits. We note that an $(\alpha, a)$-block-encoding of a matrix $\boldsymbol{A}$ can be restated, without loss of generality, as an $(\alpha\|\boldsymbol{A}\|, a)$-block-encoding of $\boldsymbol{A}/\|\boldsymbol{A}\|$.

A matrix-vector product $\boldsymbol{Ab}$ may be obtained as:

$$\boldsymbol{U_A}|0^a\rangle|\boldsymbol{b}\rangle = \begin{pmatrix} \boldsymbol{A}/\alpha & * \\ * & * \end{pmatrix} \begin{pmatrix} |\boldsymbol{b}\rangle \\ 0 \end{pmatrix} = \begin{pmatrix} \frac{1}{\alpha}\boldsymbol{A}|\boldsymbol{b}\rangle \\ * \end{pmatrix} = \frac{1}{\alpha}|0^a\rangle|\boldsymbol{Ab}\rangle + |\perp\rangle \qquad (42)$$

Measuring the ancilla qubits in the state $|0^a\rangle$ indicates a successful matrix-vector multiplication, and has a success probability of (Lin, 2022)

$$p(0^a) = \frac{1}{\alpha^2}\|\boldsymbol{A}|\boldsymbol{b}\rangle\|^2 = \frac{1}{\alpha^2}\|\boldsymbol{A}|\boldsymbol{b}\rangle\|^2 \qquad (43)$$

Block-encoded unitaries can also be used to apply a product of several matrices. We restate here the results of (Gilyén et al., 2019) for a block-encoding of a product of two matrices and generalize it to a product of several matrices.

**Lemma 1**: If $\boldsymbol{U_A}$ is an $(\alpha, a)$-block-encoding of $\boldsymbol{A}$ and $\boldsymbol{U_B}$ is a $(\beta, b)$-block-encoding of $\boldsymbol{B}$, then $(\boldsymbol{I}_b \otimes \boldsymbol{U_A})(\boldsymbol{I}_a \otimes \boldsymbol{U_B})$ is an $(\alpha\beta, a+b)$-block-encoding of $\boldsymbol{AB}$.

This can be trivially extended to products of multiple matrices as follows:

**Corollary 2**: If $\boldsymbol{U}_{\boldsymbol{A}_0}, \dots, \boldsymbol{U}_{\boldsymbol{A}_j}$ are $(\alpha_0, a_0), \dots, (\alpha_j, a_j)$-block-encodings of $\boldsymbol{A}_0, \dots, \boldsymbol{A}_j$, then $\prod_{k=j}^{0}\left[\left[\bigotimes_{\substack{0\le l\le j\\ l\neq k}} \boldsymbol{I}_l\right]\boldsymbol{U}_k\right]$ is a $\left(\prod_{k=j}^{0}\alpha_k, \sum_{0\le k\le j} a_k\right)$-block-encoding of $\prod_{k=j}^{0}\boldsymbol{A}_k$.

Naïve application of products of matrices using Corollary 2 requires a large number of qubits. The number of ancilla qubits can be reduced using a 'compression gadget' developed by (Low & Wiebe, 2018).

**Lemma 3**: (Low & Wiebe, 2018) Given unitaries $\boldsymbol{U}_{\boldsymbol{A}_k} \forall\, k \in [0, j]$, each of which is an $(\alpha_k, a_k)$-block-encoding of $\boldsymbol{A}_k$. Then we can construct an $(\alpha_{comp}, a_{comp})$-block-encoding of $\prod_{k=j}^{0}\boldsymbol{A}_k$ where $\alpha_{comp} = \prod_{k=j}^{0}\alpha_k$, $a_{comp} = \max_k a_k + \lceil \log_2 j \rceil + 1$ using each $\boldsymbol{U}_{\boldsymbol{A}_k}$ once.

For the purposes of our algorithm, we assume access to oracles that block-encode the matrix operators described in Section 3.

# 5. Quantum multigrid (qMG) algorithm

Now we state an algorithm that produces a quantum state encoding of the iterates of the multigrid method. The algorithm follows the same sequence of operations as a classical multigrid algorithm, with the difference being that the operations are performed on a block-encoded quantum state.

Towards the end of the algorithm, we obtain the state $|0\rangle^{\otimes q}|\boldsymbol{x}_i\rangle + |\perp\rangle$. Measuring the ancilla qubits in the state $|0\rangle^{\otimes q}$ and the indexing qubits in the state $|i\rangle\ \forall\, T \le i \le T + c$ indicates that the remaining qubits encode the desired final iterate $|\boldsymbol{v}_{\mathcal{V},0,2\nu-1}\rangle$ of the multigrid algorithm.

Algorithm 2. Quantum Multigrid Algorithm

**Inputs**:
$O_{\boldsymbol{v}_{in}}, O_{\boldsymbol{f}}, O_{\boldsymbol{Pre}_{V,L,v}}, O_{\boldsymbol{Post}_{V,L,v}}, O_{\boldsymbol{Res}_{V,L}} O_{\boldsymbol{Pro}_{V,L}}, O_{\boldsymbol{PreCopy}_{V,L,v}}, O_{\boldsymbol{PostCopy}_{V,L,v}}, O_{\boldsymbol{Copy}_j}$
$\mathcal{V}, \mathcal{L}, \nu, p$
$|0\rangle^{\otimes q+m}$

**Output**: $|\boldsymbol{v}_{\mathcal{V},0,2\nu-1}\rangle$

Prepare $|0\rangle^{\otimes q}|\boldsymbol{x}_{in}\rangle$ using oracles $O_{\boldsymbol{v}_{in}}, O_{\boldsymbol{f}}$
**For** $V = 0 : \mathcal{V}$
  **For** $L = 0 : \mathcal{L}$
    **For** $v = 1 : \nu$
      Apply pre-smoothing or calculate residual using $O_{\boldsymbol{Pre}_{V,L,v}}$
    Restrict residual using $O_{\boldsymbol{Res}_{V,L}}$
    **For** $v = 1 : \nu, \nu + 1 : 2\nu - 1$
      Copy residual using $O_{\boldsymbol{PreCopy}_{V,L,v}}, O_{\boldsymbol{PostCopy}_{V,L,v}}$
  **For** $L = \mathcal{L} : 0$
    **For** $v = \nu + 1 : 2\nu - 1$
      Apply post-smoothing using $O_{\boldsymbol{Post}_{V,L,v}}$
    Prolong solution using $O_{\boldsymbol{Pro}_{V,L}}$
**For** $j = 1 : c$

|     Copy final iterate using $O_{\boldsymbol{Copy}_j}$
Measure ancilla qubits $|\psi_a\rangle$ and indexing qubits $|\psi_i\rangle$
**If** $|\psi_a\rangle! = |0\rangle^{\otimes q}$ or $|\psi_i\rangle! = |i\rangle \; \forall \, T \leq i \leq T + c$
|     Restart
**End**

We now analyze the complexity of the individual components of the algorithm which include state preparation, success probabilities, and number of matrix multiplications. We arrive at the overall complexity in Section 5.3.

## 5.1. State Preparation

At the beginning of the algorithm, we need to prepare the initial state $|\boldsymbol{x}_{in}\rangle$. This can be done efficiently using the following lemma by noting that $\boldsymbol{x}_{in}$ contains $\mathcal{V}(\nu - 1)^2$ copies of $\boldsymbol{f}$.

**Lemma 4:** (Berry et al., 2017) Let $\mathcal{O}_{\boldsymbol{v}_{in}}$ be a unitary that maps $|1\rangle|\psi\rangle$ to $|1\rangle|\psi\rangle$ for any $|\psi\rangle$ and maps $|0\rangle|0\rangle$ to $|0\rangle|\boldsymbol{v}_{in}\rangle$ where $\boldsymbol{v}_{in} = \boldsymbol{v}_{in}/\|\boldsymbol{v}_{in}\|$. Let $\mathcal{O}_{\boldsymbol{f}}$ be a unitary that maps $|0\rangle|\psi\rangle$ to $|0\rangle|\psi\rangle$ for any $|\psi\rangle$ and maps $|1\rangle|0\rangle$ to $|1\rangle|\boldsymbol{f}\rangle$ where $\boldsymbol{f} = \boldsymbol{f}/\|\boldsymbol{f}\|$. Suppose we know $\|\boldsymbol{v}_{in}\|$ and $\|\boldsymbol{f}\|$. Then the state proportional to $\sum_{i=1}^{l} |i\rangle|\boldsymbol{f}\rangle + |0\rangle|\boldsymbol{v}_{in}\rangle$ can be produced with $O(1)$ calls to $\mathcal{O}_{\boldsymbol{v}_{in}}$ and $\mathcal{O}_{\boldsymbol{f}}$, and $O(\text{poly}\log \mathcal{V}\nu)$ elementary quantum gates.

We use Lemma 2 to factor in the complexity of preparing the state $|\boldsymbol{x}_{in}\rangle$.

## 5.2. Success probabilities

Since the goal of a QLSA is to produce the quantum state encoding a solution with precision $\epsilon$, in this case the final iterate $\left|\boldsymbol{v}_{\mathcal{V},0,2\nu-1}\right\rangle$, the final iterate or one of its copies needs to be extracted from the state $|\boldsymbol{x}\rangle$ by performing a measurement of the ancilla and indexing registers. Measuring these registers in the state $|0\rangle^{\otimes q}$ and $|i\rangle \; \forall \, T \leq i \leq T + c$ respectively indicates a successful measurement, after which the remaining qubits encode the desired final iterate $\left|\boldsymbol{v}_{\mathcal{V},0,2\nu-1}\right\rangle$.

In the following sections we analyze individually the complexity of successful measurements of these registers.

### 5.2.1. Ancilla registers

We first note that the algorithm proceeds by applying $T + c + 1$ of $(\zeta_i, \xi_i)$-block-encoded matrix multiplications to the prepared initial state, where $i \in T + c + 1$. Therefore, using Corollary 2 the action of the algorithm $\boldsymbol{qMG}$ can be viewed as a $(\mathrm{Z}, \Xi)$-block-encoding of the product of the matrices as

$$(\boldsymbol{qMG})|0\rangle^{\otimes \Xi}|\boldsymbol{x}_{in}\rangle = \frac{1}{\mathrm{Z}}|0\rangle^{\otimes \Xi}|\boldsymbol{x}\rangle + |\perp\rangle \qquad (44)$$

Using Lemma 3 $\Xi = \max_i \xi_i + \lceil \log_2(T + c + 1) \rceil + 1$ is the total number of ancilla qubits used for the block-encodings and the probability of successfully measuring the ancilla qubits in the state $|0\rangle^{\otimes \Xi}$ is

$$p\left(|0\rangle^{\otimes \Xi}\right) = \frac{1}{\prod_{i=T+c+1}^{0} \zeta_i^2} \frac{\left\| \prod_{i=T+c+1}^{0} Operation_i |\boldsymbol{x_{in}}\rangle \right\|^2}{\prod_{i=T+c+1}^{0} \|Operation_i\|^2} = \frac{1}{\mathrm{Z}^2} \qquad (45)$$

$\mathrm{Z}$ depends on the subnormalization factors $\zeta_i$ of the individual oracles and their implementation and the further subnormalization arising from the factor $\frac{\prod_{i=T+c+1}^{0}\|Operation_i\|}{\left\|\prod_{i=T+c+1}^{0} Operation_i\right\|}$ where $Operation_i$ denotes the block matrix corresponding to the $i^{th}$ operation in the multigrid method. Large $\zeta_i$ and $\mathrm{Z}$ can make the success probability vanishingly small. In the following section we investigate the probability of successful measurement of the index register.

#### 5.2.2. Index register

The probability of successfully measuring the index register in a state $|i\rangle\ \forall\ T \le i \le T+c$ can be controlled using the number of copy steps $c$. Since the convergence of the multigrid algorithm is monotonic, an initial guess of $\mathbf{0}$ will lead to growing solutions across V-cycles. The solution at coarser scales is a growing solution since the initial guess for the correction is $\mathbf{0}$. Using these arguments, we see that for an initial guess of $\mathbf{0}$ the norm of the final iterate will be greater than the previous iterate. Furthermore, if an initial guess is provided with an error $\epsilon$, we can place a bound on the norm of the final iterate due to monotonic convergence. Therefore, we may use the following lemmas from (Raisuddin & De, 2023) to select an appropriate $p$.

**Lemma 5**: Given a monotonically convergent scheme and starting with an initial guess of $\mathbf{0}$ with a choice of $c = T - 1$, successfully measuring the first register $|i\rangle$ in the state such that $i \in [l, l+c]$ has probability $p \ge 1/2$.

**Lemma 6**: Given a monotonically convergent scheme starting with an initial guess $\boldsymbol{x}_{in}$ and error $\epsilon = \|\widetilde{\boldsymbol{x}} - \boldsymbol{x}_{in}\| \le \|\widetilde{\boldsymbol{x}}\|$ , where $\widetilde{\boldsymbol{x}} = \boldsymbol{A}^{-1}\boldsymbol{b}$ , and a choice of $c = T - 1$, the probability $p$ of successfully measuring the first register $|i\rangle$ in the state such that $i \in [T, T+c]$ is $p \ge \frac{1}{2}\left(\frac{\|\widetilde{\boldsymbol{x}}\|-\epsilon}{\|\widetilde{\boldsymbol{x}}\|+\epsilon}\right)^2$.

We note these are conservative estimates for $c$.

### 5.3. Complexity analysis

We now analyze the complexity of the quantum multigrid algorithm in resources and runtime.

**Theorem 7**: Given oracles $O_{\boldsymbol{v}_{in}}, O_{\boldsymbol{f}}, O_{\boldsymbol{Pre}_{V,L,v}}, O_{\boldsymbol{Post}_{V,L,v}}, O_{\boldsymbol{Res}_{V,L}}, O_{\boldsymbol{Pro}_{V,L}}, O_{\boldsymbol{PreCopy}_{V,L,v}}, O_{\boldsymbol{PostCopy}_{V,L,v}}$ and $O_{\boldsymbol{Copy}_j}$, the quantum multigrid algorithm can prepare the state

$$|\psi\rangle = |0\rangle^{\otimes m}|\boldsymbol{x}\rangle + |\perp\rangle \tag{46}$$

in $O\left(\text{poly}\log\frac{N}{\tilde{\epsilon}}\right)$ time (accesses to oracles) and requires $O\left(\text{poly}\log\frac{N}{\tilde{\epsilon}}\right)$ qubits. The success probability of measuring the ancilla register and index register in a desired state $|0\rangle^{\otimes m}|i\rangle\ \forall\ i \in [T, 2T]$ is

$$p = O\left(\frac{1}{\mathrm{Z}^2}\right) \tag{47}$$

The success probability can be boosted to $O(1)$ using $O(\mathrm{Z})$ rounds of amplitude amplification, leading to an overall runtime of $O\left(\mathrm{Z}\frac{\mathrm{Z}}{\||\boldsymbol{x}\rangle\|}\text{poly}\log\frac{N}{\tilde{\epsilon}}\right)$ using $O\left(\text{poly}\log\frac{N}{\tilde{\epsilon}}\right)$ qubits.

**Proof 5**: First, we note that solution error decreases exponentially with the number of V-cycles, i.e., $\mathcal{V} = O\left(\log\left(\frac{\|\boldsymbol{e}\|}{\|\boldsymbol{e}_0\|}\right)^{-1}\right) = O\left(\log\left(\frac{\epsilon}{\epsilon_0}\right)^{-1}\right) = O(\log(\tilde{\epsilon})^{-1})$. We also note that the number of grid levels $\mathcal{L}$ scales logarithmically with the number of unknowns, i.e., $\mathcal{L} = O(\log N)$ where $N$ is the number of unknowns at the finest grid level. Finally, we note that the number of smoothing steps remains constant regardless of the problem size, i.e., $\nu = O(1)$.

This indicates that the algorithm requires $O\left(\operatorname{poly}\log\frac{N}{\tilde{\epsilon}}\right)$ smoothing steps and residual copy steps, and $O\left(\operatorname{poly}\log\frac{N}{\tilde{\epsilon}}\right)$ restriction, prolongation, and residual calculation steps. The final copying step requires $c$ applications of $O_{\boldsymbol{Copy}_j}$. Using an initial guess of $\mathbf{0}$ and Lemma 4, we can choose $c = T - 1 = O\left(\operatorname{poly}\log\frac{N}{\tilde{\epsilon}}\right)$ for an $O(1)$ success probability of successful measurement of the index register.

Therefore, the total number of block-matrix products scales as $O\left(\operatorname{poly}\log\frac{N}{\tilde{\epsilon}}\right)$. Since the application of the matrix products requires block-encodings with $\Xi = \max_i \xi_i + \left\lceil \log_2\left(\operatorname{poly}\log\frac{N}{\tilde{\epsilon}}\right)\right\rceil + 1$ ancilla qubits, in addition to $\log_2 T + c = O\left(\operatorname{poly}\log\frac{N}{\tilde{\epsilon}}\right)$ indexing qubits and $\log_2 N$ work register qubits we require an additional $O(T + c)$ ancilla qubits for an overall $O\left(\operatorname{poly}\log\frac{N}{\tilde{\epsilon}}\right)$ total number of qubits, assuming the block-encodings of the matrices are efficient, i.e. require at most $O\left(\operatorname{poly}\log\frac{N}{\tilde{\epsilon}}\right)$ ancilla qubits.

Therefore, to prepare the state $|0\rangle^{\otimes q+m}|\boldsymbol{x}\rangle + |\perp\rangle$ we require $O\left(\operatorname{poly}\log\frac{N}{\tilde{\epsilon}}\right)$ queries to the oracles and $O\left(\operatorname{poly}\log\frac{N}{\tilde{\epsilon}}\right)$ qubits.

The success probability of measuring the ancilla qubits in the state $|0\rangle^{\otimes q+m}$ is $\frac{1}{Z^2}$. Since the probability of successfully measuring the index register in the quantum state $|\boldsymbol{x}\rangle$ is $p = \frac{1}{2} = O(1)$, the combined probability of obtaining a quantum state of the final iterate $|\boldsymbol{x}_{out}\rangle$ by measuring the ancilla and index qubits is $O\left(\frac{1}{Z^2}\right)$.

Amplitude amplification can be used to improve the overall complexity of quantum algorithms by amplifying an $O(p)$ success probability of the desired output to $O(1)$ by repeating the algorithm $O(\sqrt{p})$ times, leading to a quadratic improvement in complexity (Brassard et al., 2002). Using this result, the success probability of obtaining the final iterate $|\boldsymbol{x}_{out}\rangle$ can be boosted to $O(1)$ using $O(Z)$ rounds of amplitude amplification. Combining these we arrive at an overall complexity of $O\left(Z\operatorname{poly}\log\frac{N}{\tilde{\epsilon}}\right)$ oracle queries and $O\left(\operatorname{poly}\log\frac{N}{\tilde{\epsilon}}\right)$ qubits. □

# 6. Numerical examples

We investigate the qMG algorithm using problems in one- and two- dimensions to investigate the qubit requirements and behavior of the success probability $p$ for various problem sizes. We use MATLAB to create $\boldsymbol{x}_{in}$ and apply the sequence of operations in qMG to produce the vector $\boldsymbol{x}$.

## 6.1. 1D problem

We consider Poisson's problem in 1D with homogeneous Dirichlet boundary conditions and a combination of Dirichlet and Neumann boundary conditions.

*Find* $u: \bar{\Omega} \to \mathbb{R}$ *such that*

$u"(x) - 1 = 0 \; in \; \Omega \in (0,1)$

subject to

Case 1: $u(0) = u(1) = 0$

Case 2: $u(0) = 0 \, , u'(1) = 1$

Piecewise linear finite elements are used to discretize the problem over regular intervals. Various mesh sizes are considered, with the parameters of the finest meshes for Cases 1 and 2 shown in Table. Figure 1 shows the fully converged multigrid solution, the exact solution, and the solution error for Cases 1 and 2.

*Table1 Number of grid levels $\mathcal{L}$, V-cycles $\mathcal{V}$, smoothing steps $\nu$, unknowns $N$, and length of vector x for the finest grid for 1D Cases 1 and 2.*

| Case | $\mathcal{L}$ | $\mathcal{V}$ | $\nu$ | $\tilde{\epsilon}$ | $N$ | $len\,(\boldsymbol{x})$ | $\lceil \log_2 N \rceil$ | $\lceil \log_2 len\,(\boldsymbol{x}) \rceil$ |
|---|---|---|---|---|---|---|---|---|
| 1 | 12 | 15 | 6 | $10^{-10}$ | 8191 | 46926239 | 13 | 26 |
| 2 | 13 | 15 | 6 | $10^{-10}$ | 8192 | 50601984 | 13 | 26 |

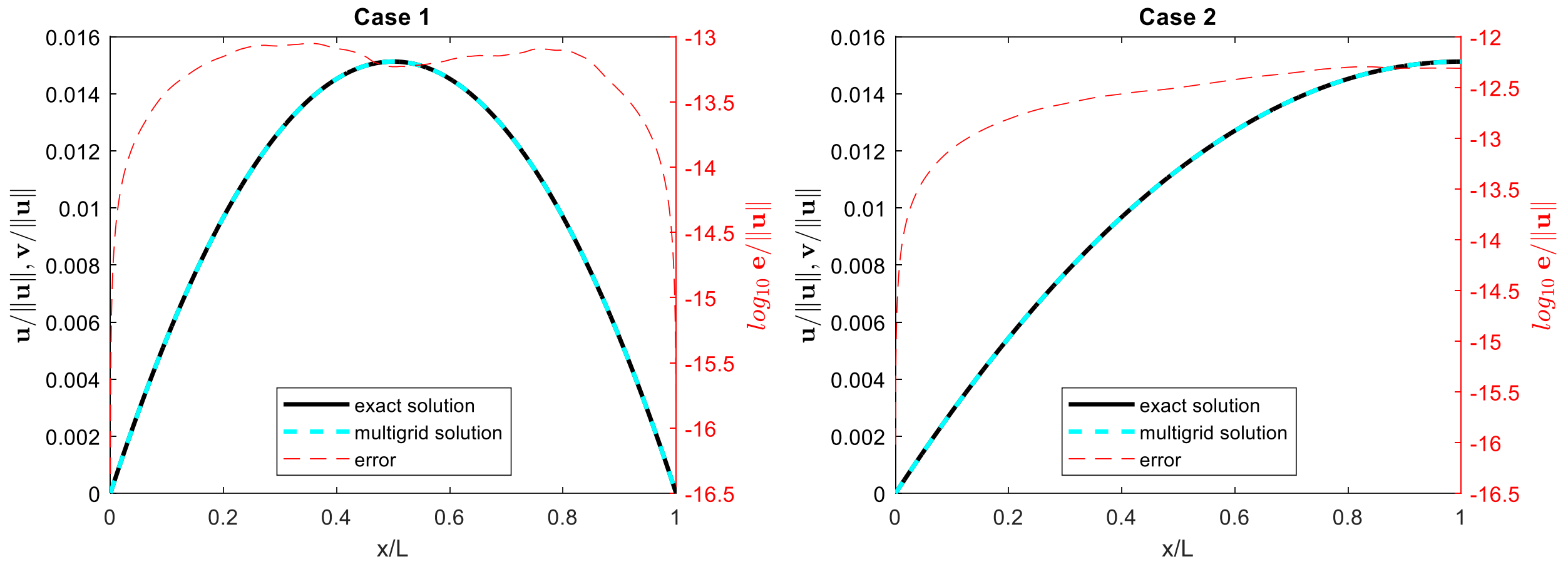

*Figure 1 Fully converged and exact solutions $\boldsymbol{u}/\|\boldsymbol{u}\|$ and $\boldsymbol{v}/\|\boldsymbol{u}\|$ with the error $\boldsymbol{e}/\|\boldsymbol{u}\|$ at the finest grid for Cases 1 and 2.*

Figure 2 demonstrates the number of additional qubits required to store $\boldsymbol{x}$ for a solution when reducing the relative error $\tilde{\epsilon}$ by a factor of $10^{-10}$ for various problem sizes $N$, as a multiple of the $\log N$ number of qubits needed to store $\boldsymbol{v}$. This is calculated by using the number of V-cycles observed experimentally to obtain $\tilde{\epsilon} \leq 10^{-10}$, as seen in Figure 3, to obtain $len\,(\boldsymbol{x}) = N(T + c + 1)$ using Equation 7 and the choice $c = T - 1$.

We see that as the problem size increases, the factor $\frac{\log len\,(\boldsymbol{x})}{\log N}$ decreases steadily below 2, meaning that the number of additional qubits to represent $\boldsymbol{x}$ is indeed $O(\text{poly} \log N)$ when $\tilde{\epsilon}$ is kept constant.

However, polynomial growth in $\log N$ is a conservative estimate and we see that for large $N$ the number of additional qubits to represent $\boldsymbol{x}$ approaches $O(\log N)$.

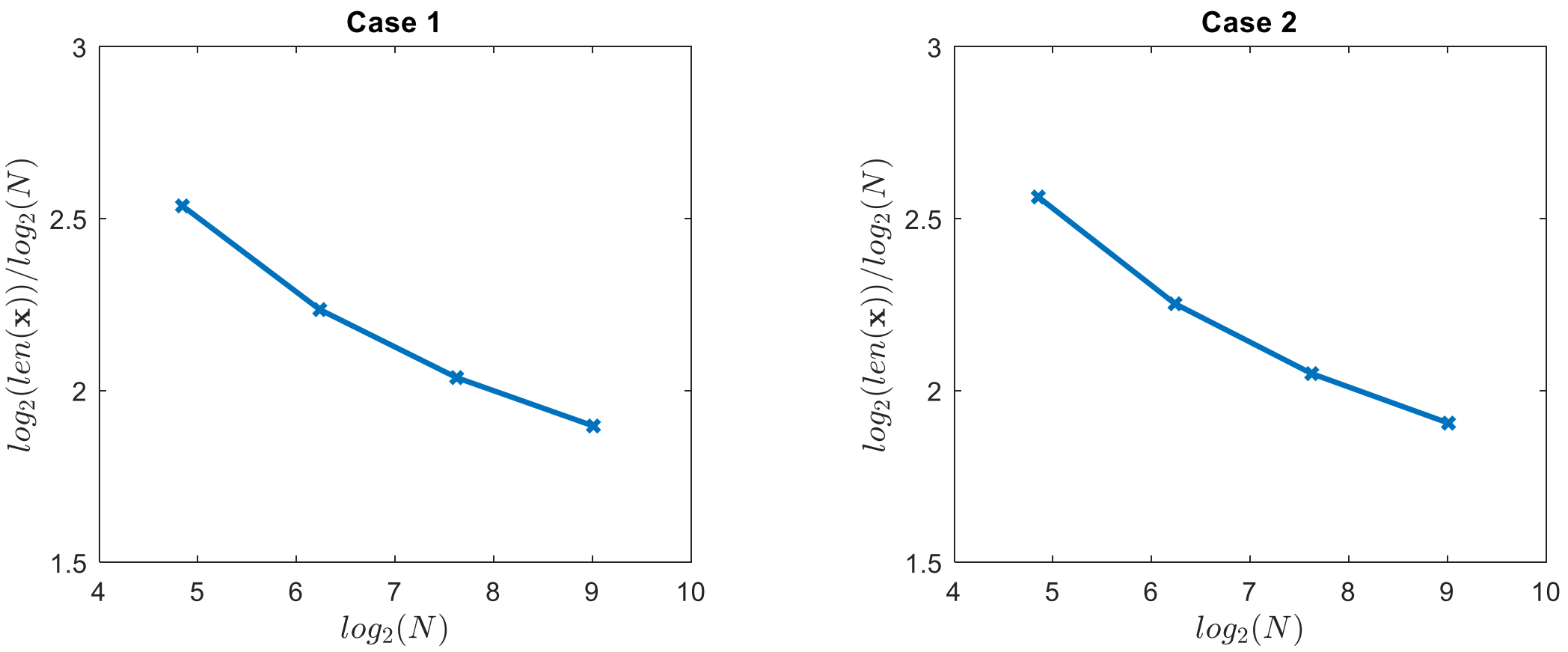

*Figure 2 Multiple of additional qubits required for larger problems for 1D Cases 1 and 2 to obtain* $\tilde{\epsilon} \leq 10^{-10}$

In Figure 3 we plot the convergence of the multigrid V-cycles, the norms of the blocks of $\boldsymbol{x}$ as a ratio of the block $\|\boldsymbol{x}_i\|$ to the desired output block $\|\boldsymbol{x}_{out}\|$, and the success probabilities of successfully measuring the index qubit for various $N$ and number of V-cycles. As predicted, the norms of the undesired blocks are mostly small compared to the desired final iterate. This is reflected in the probabilities of successfully measuring the index qubit being much larger than the predicted $p \geq 0.5$. As the number of V-cycles increases, the success probabilities decrease slightly, with an almost constant success probability seen for many V-cycles. Similarly, as the problem size is increased, the success probability also increases. This indicates that for larger problems, successful measurement of the index register is more likely.

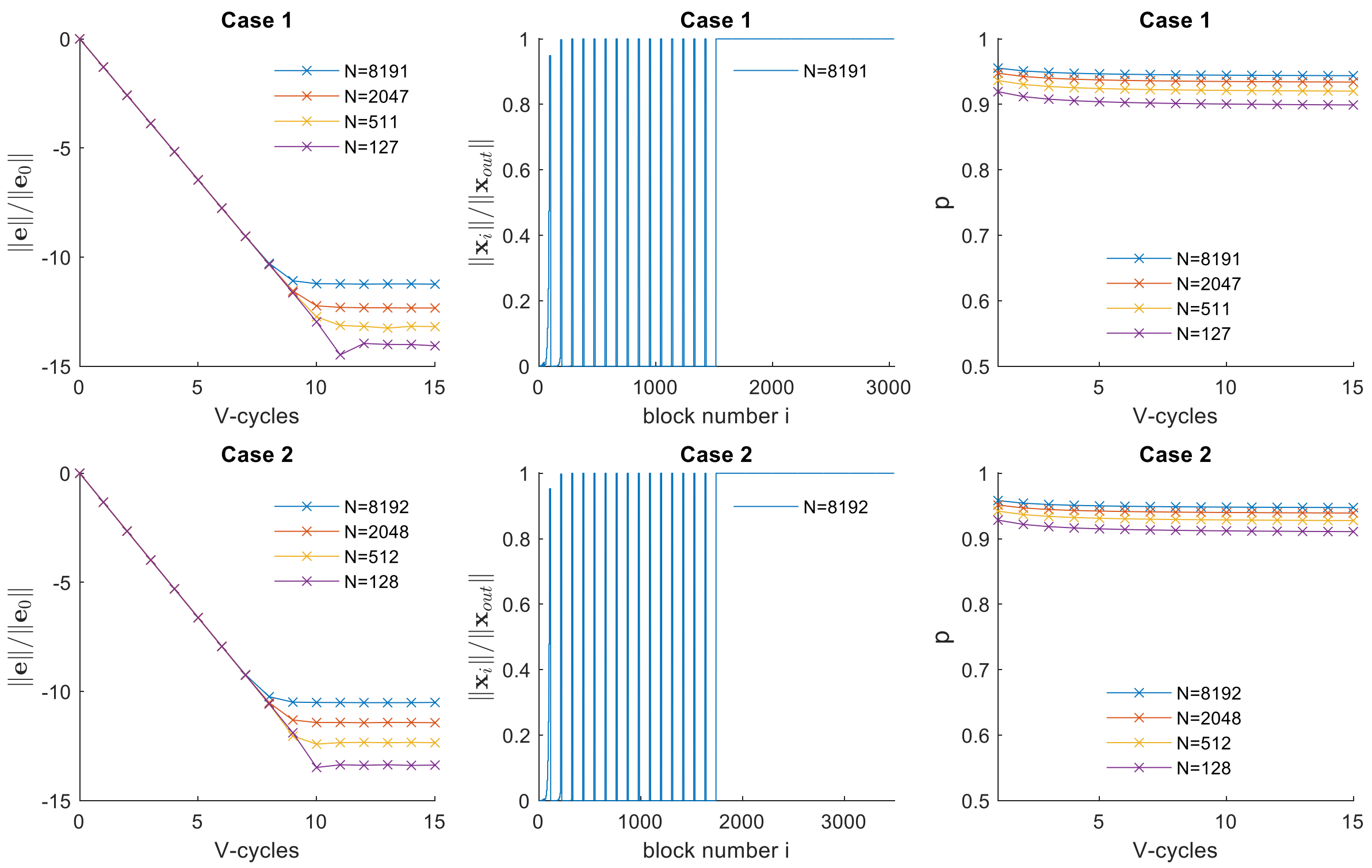

*Figure 3 Convergence of multigrid V-cycles for various problem sizes, the norms of the blocks as a ratio of the norm of the final iterate, and the success probability p of measuring the index register for various problem sizes and number of V-cycles for 1D Cases 1 and 2.*

## 6.2. 2D problem

We consider Poisson's problem on a square domain:

*Find* $u: \bar{\Omega}^2 \rightarrow \mathbb{R}$ *such that*

$$\nabla^2 u - f = 0 \; in \; \Omega \in (0, L) \times (0, L)$$

subject to boundary conditions

$$u = g \; on \; \Gamma_g$$

$$-q_i n_i = h \; on \; \Gamma_h$$

where $q_i n_i$ is the unit outward normal to $\Gamma = \Gamma_g \cup \Gamma_h$ and $\Gamma_g \cap \Gamma_h = \emptyset$.

We consider the cases presented in Table 1 for $\Gamma, g, h$, and $f$.

*Table 1 Boundary conditions for 2D Cases 1-5*

| | $\Gamma_g$ | $g$ | $\Gamma_h$ | $h$ | $f$ |
|---|---|---|---|---|---|
| Case 1 | $(0,y),(L,y),(x,0),(x,L)$ $\forall x,y \in [0,L]$ | 0 | $\emptyset$ | $-$ | 1 |
| Case 2 | $(0,y),(L,y),(x,0)$ $\forall x,y \in [0,L]$ | 0 | $(x,L)$ $\forall x \in [0,L]$ | 0 | 1 |
| Case 3 | $(0,y),(L,y)$ $\forall y \in [0,L]$ | 0 | $(x,0),(x,L)$ $\forall x \in [0,L]$ | 0 | 1 |
| Case 4 | $(0,y),(x,0)$ $\forall x,y \in [0,L]$ | 0 | $(L,y),(x,L)$ $\forall x,y \in [0,L]$ | 0 | 1 |
| Case 5 | $(0,y)$ $\forall y \in [0,L]$ | 0 | $(L,y),(x,0),(x,L)$ $\forall x,y \in [0,L]$ | 0 | 1 |

Four-noded bilinear finite elements are used to discretize the problem over regular grids. We discretize over various grid sizes, with the finest grid sizes shown in Table 2 for Cases 1-5. Figure 2 shows the fully converged multigrid solution for Cases 1-5.

*Table 2 Number of grid levels $\mathcal{L}$, V-cycles $\mathcal{V}$, smoothing steps $\nu$, grid points along $x$ and y directions $N_x$, $N_y$, problem size $N$, and length of $x$ for the finest grid size for 2D Cases 1-5.*

| Case | $\mathcal{L}$ | $\mathcal{V}$ | $\nu$ | $N_x$ | $N_y$ | $N$ | $\tilde{\epsilon}$ | $len\ (\boldsymbol{x})$ | $\lceil \log_2 N \rceil$ | $\lceil \log_2 len\ (\boldsymbol{x}) \rceil$ |
|---|---|---|---|---|---|---|---|---|---|---|
| 1 | 6 | 25 | 6 | 127 | 127 | 16129 | $10^{-10}$ | 79693389 | 14 | 27 |
| 2 | 6 | 35 | 6 | 127 | 64 | 8128 | $10^{-10}$ | 55603648 | 14 | 26 |
| 3 | 6 | 45 | 6 | 127 | 65 | 8255 | $10^{-10}$ | 72156955 | 13 | 27 |
| 4 | 6 | 45 | 6 | 64 | 64 | 4096 | $10^{-10}$ | 35803136 | 12 | 26 |
| 5 | 6 | 50 | 6 | 64 | 65 | 4160 | $10^{-10}$ | 36362560 | 13 | 26 |

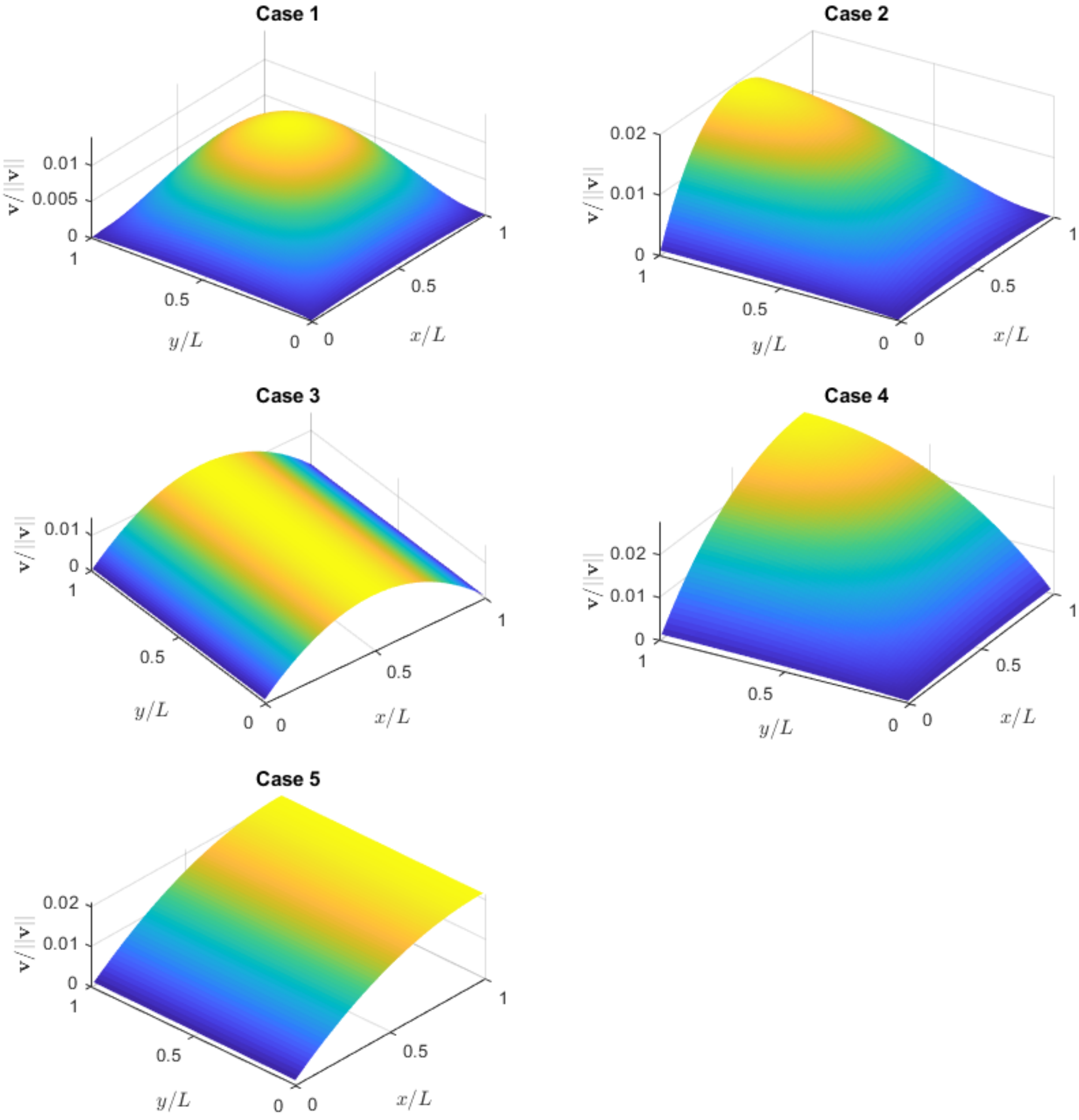

*Figure 4 Fully converged solutions $\boldsymbol{v}/||\boldsymbol{v}||$ for the finest grid sizes for 2D Cases 1-5*

Figure 5 shows the number of additional qubits required to store $\boldsymbol{x}$ compared to the number of qubits required to store $\boldsymbol{v}$ as a multiple of $\log N$ when the error is reduced by a factor of $10^{-10}$. Similar to the 2D case, we see that the overhead in additional qubits decreases for larger problems, demonstrating that $O(\text{poly} \log N)$ is a conservative estimate of the number of qubits needed to store $\boldsymbol{x}$, since it approaches $O(\log N)$.

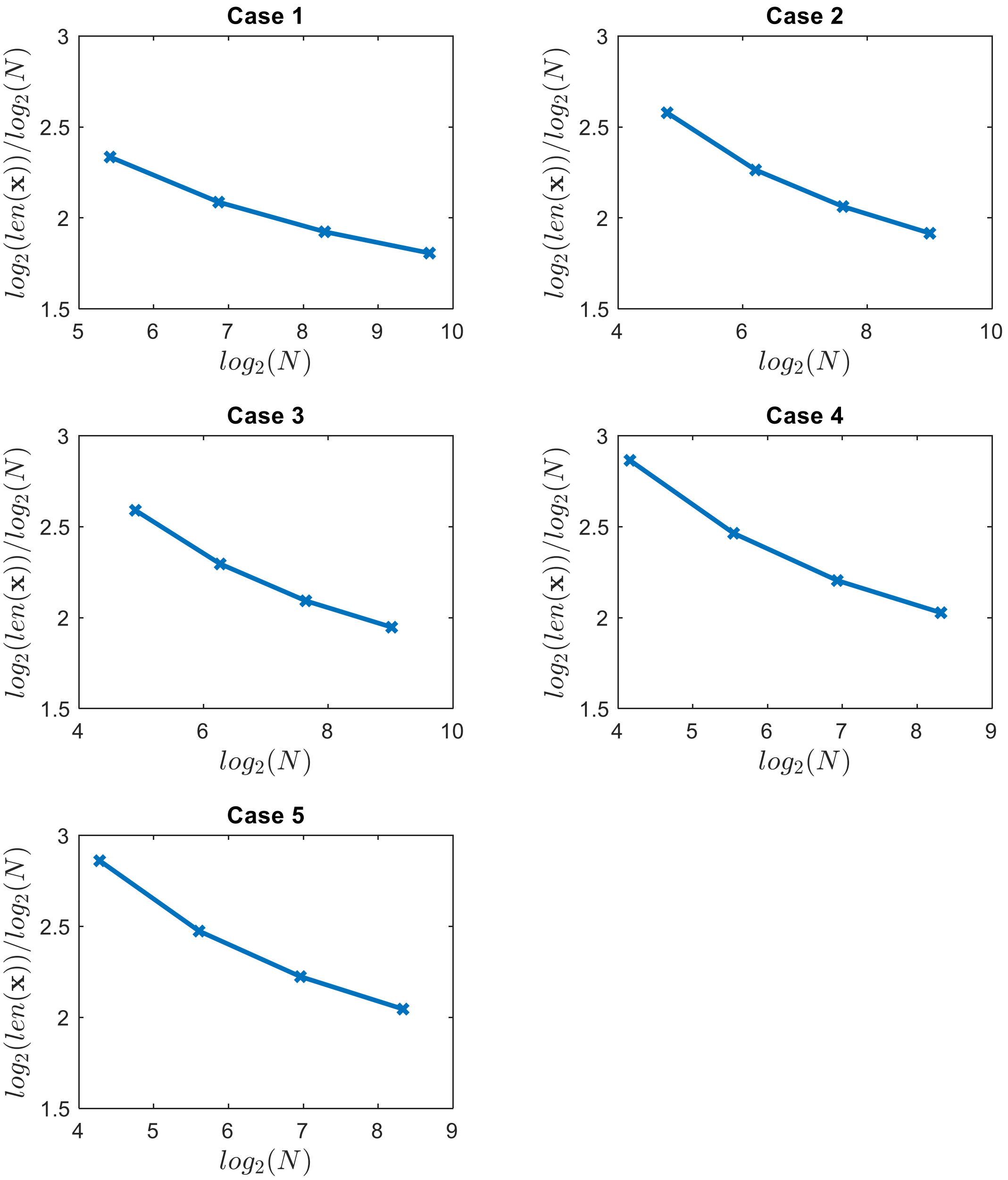

*Figure 5 Multiple of additional qubits with increasing problem sizes for 2D Cases 1-5 to obtain* $\tilde{\epsilon} \leq 10^{-10}$

In Figure 6 we show the convergence of the multigrid V-cycles, the ratios of the blocks $\|\boldsymbol{x}_i\|/\|\boldsymbol{x}_{out}\|$, and the success probability $p$ of successfully measuring the indexing register. Once again, we see that the theoretical bound of $p \geq 0.5$ is conservative, with actual values converging to $p > 0.9$ for large $N$ and a large number of V-cycles. This indicates that $p$ scales favorably for larger problems.

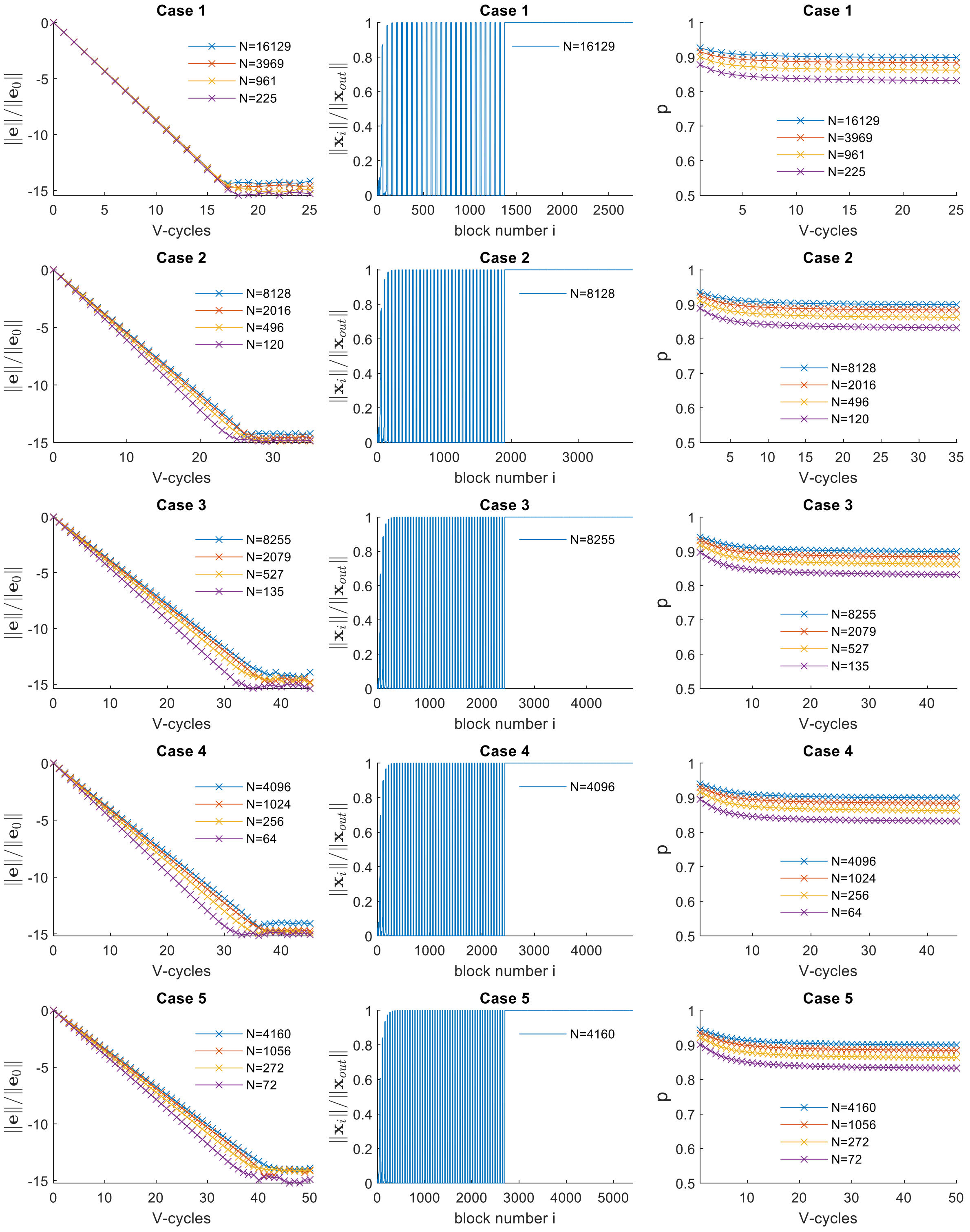

*Figure 6 Convergence of multigrid V-cycles for various problem sizes, the norms of the blocks as a ratio of the norm of the final iterate, and the probability $p$ of measuring the index register for various problem sizes and number of V-cycles for 2D Cases 1-5.*

# 7. Conclusion

In this work, we introduce qMG, an algorithm designed to execute V-cycle multigrid operations on gate-based quantum computing architectures. The algorithm encodes the series of linear operations associated with the V-cycle into large matrices, while the corresponding multigrid iterates are encoded into an extensive vector. A quantum state, representing the sequence of multigrid iterates, is constructed with exponential efficiency. The extraction of the final iterate is performed by measuring a specific block-indexing register. To ensure efficient extraction, the final iterate is copied to subsequent blocks, resulting in an $O(1)$ probability of successful measurement.

We show numerically that the success probability of extracting the final iterate from the sequence of iterates is much larger than the theoretically derived bound and that it further improves for larger problem sizes and the number of V-cycles. We also demonstrate the scaling of the qubit overheads, showing that the number of qubits required to store the vector $\boldsymbol{x}$ does not grow significantly with $N$.

Although the state encoding the sequence of iterates $\boldsymbol{x}$ can be produced exponentially efficiently as a subspace of the quantum registers, the success probability of measuring the ancilla qubits in the desired state can be low due to the sub-normalization of the block encodings. The values of the sub-normalization constants depend on the specific implementation of the oracles and the norm of the products of matrices. This emphasizes the need to design oracles that are not only efficient in the number of gates and ancilla qubits but also in the sub-normalization constants. A similar issue is reported in (Shao & Xiang, 2018).

Recent work has shown that for a product of matrices the subnormalization factor $\alpha_i$ of each individual block-encoding can be reduced to $\approx 1$ by amplifying each block using the Uniform Singular Value Amplification (Gilyén et al., 2019) method. This comes at an expense of $\approx O\left(\alpha_i \log\frac{\alpha_i}{\epsilon_i}\right)$ accesses to each block encoding and an approximation error $\epsilon_i$ in the 2-norm and can be performed efficiently using quantum signal processing (Fang et al., 2023). This converts the accumulation of computational cost arising from subnormalization factors from a product of $\alpha_i$ into (approximately) a sum of $\alpha_i$, which is an exponential improvement. However, if $\frac{\Pi\|A_i\|}{\|\Pi A_i\|}$ is large, the success probability remains small. Improved techniques to apply products of matrices can enable qMG to improve its success probability for a fully exponential speedup.

In our analysis, we choose a conservative value for the number of copies $c$. A tighter bound will not yield a substantial improvement since the contribution to the overall complexity scales the same as the complexity of the multigrid operations.

Although the presented method realizes a series of V-cycles, it can be extended to include other multigrid cycles like the W-cycle and the F-cycle, which require a more complex sequence or operations for better convergence (Mandel & Parter, 1990). Furthermore, the method applies to both geometric and algebraic multigrid approaches.

Domain decomposition methods are a similar iterative approach that applies relaxation operators to alternating regions of the domain. The method outlined in this paper can be extended to realize a quantum domain decomposition approach for similar scaling. However, the accumulation of sub-normalization factors leading to low success probabilities is not expected to be resolved using a domain decomposition approach.

Although the algorithm is readily posed as a variable-time stopping algorithm by measuring the ancilla qubits between matrix multiplication to check for success, variable-time amplitude amplification (Ambainis, 2012) or efficient variable-time amplitude amplification (Chakraborty et al., 2019) cannot boost the success probability of the method beyond simple amplitude amplification since the average stopping time is not less than the maximum stopping time.

## Acknowledgements

The authors would like to acknowledge the support of this work by NIBIB R01EB0005807, R01EB25241, R01EB033674 and R01EB032820 grants.